\documentclass[aps,prl,twocolumn,superscriptaddress,showpacs]{revtex4}
\usepackage{graphicx}
\usepackage{epsfig}
\usepackage{latexsym}
\usepackage[dvips]{color}

\begin{document}


\title{Inhomogeneous level splitting in Pr$_{2-x}$Bi$_{x}$Ru$_2$O$_7$}

\author{J. van Duijn}
\affiliation{Department of Physics and Astronomy, Johns Hopkins University, Baltimore, Maryland 21218}
\affiliation{ISIS Facility, Rutherford Appleton Laboratory, Chilton, Didcot, OX11 0QX, U.K.}
\author{K. H. Kim}
\altaffiliation[Permanent address:]{ CSCMR \& School of Physics, Seoul National University, Seoul 151-747, S. Korea}
\affiliation{MST-NHMFL, Los Alamos National Laboratory, NM 87544, USA}
\author{N. Hur}
\affiliation{Department of Physics and Astronomy, Rutgers University, Piscataway, New Jersey 08854}
\author{D. Adroja}
\affiliation{ISIS Facility, Rutherford Appleton Laboratory, Chilton, Didcot, OX11 0QX, U.K.}
\author{M. A. Adams}
\affiliation{ISIS Facility, Rutherford Appleton Laboratory, Chilton, Didcot, OX11 0QX, U.K.}
\author{Q. Z. Huang}
\affiliation{NIST Centre for Neutron Research, National Institute of Standards and Technology, Gaithersburg, Maryland 20899}
\author{M. Jaime}
\affiliation{MST-NHMFL, Los Alamos National Laboratory, NM 87544, USA}
\author{S. -W. Cheong}
\affiliation{Department of Physics and Astronomy, Rutgers University, Piscataway, New Jersey 08854}
\author{C. Broholm}
\affiliation{Department of Physics and Astronomy, Johns Hopkins University, Baltimore, Maryland 21218}
\affiliation{NIST Centre for Neutron Research, National Institute of Standards and Technology, Gaithersburg, Maryland 20899}
\author{T. G. Perring}
\affiliation{ISIS Facility, Rutherford Appleton Laboratory, Chilton, Didcot, OX11 0QX, U.K.}
\date{\today}

\begin{abstract}
We report that Bi doping drives Pr$_{2-x}$Bi$_{x}$Ru$_2$O$_7$ from an antiferromagnetic insulator ($x=0$) to a
metallic paramagnet ($x\approx 1$) with a broad low $T$ maximum in $C/T$. Neutron scattering
reveals local low energy spin excitations ($\hbar\omega\approx 1$ meV) with a spectrum that is
unaffected by heating to $k_BT>>\hbar\omega$. We show that a continuous distribution of splittings of the non-Kramers
Pr$^{3+}$ ground state doublet such as might result from various types of lattice strain can account for all the data.
\end{abstract}


\maketitle
Rare earth ions  with an even number of electrons in a partially filled 4f-shell require a highly symmetric  environment to retain ground state degeneracy. Such non-Kramers degeneracy is consequently unstable towards lattice distortions, which can occur spontaneously through the Jahn-Teller effect, dynamically through interactions with phonons,~\cite{boothroyd} or at random due to static inhomogeneities~\cite{carini}. In this letter we describe an inadvertent encounter with the latter effect while exploring the unusual thermo-magnetic properties of a diluted frustrated magnet,  Pr$_{2-x}$Bi$_{x}$Ru$_2$O$_7$. While the enhanced specific heat of the material is reminiscent of ultra heavy fermion behavior, the temperature ($T$) and wave vector ($Q$) independent neutron spectra and the diverging susceptibility implicate inhomogeneous level splitting of the non-Kramers Pr doublet. Our comprehensive analysis links neutron, specific heat, and susceptibility data and is important because similar physics may be operative in other magnets of recent notoriety.

The parent material, Pr$_2$Ru$_2$O$_7$, contains two sub-lattices of magnetic ions on the vertices of corner-sharing tetrahedra. This lattice is notable because antiferromagnetic nearest neighbor interactions do not select a unique long range ordered spin configuration but a highly degenerate manifold. While  Pr$_2$Ru$_2$O$_7$ does develop magnetic order at $T_N=165$ K, dilution of the Pr$^{3+}$ sites by the smaller isovalent Bi$^{3+}$ ion converts the ordered insulator to a metallic  paramagnet. Similar behavior was found in other rare earth ruthenates, however, the Pr based material has a greatly enhanced low$-T$ specific heat with analogies to non-fermi-liquid and heavy fermion systems. Here we link the anomalies to static inhomogeneous splitting of the non-Kramers Pr$^{3+}$ doublet, and consider various options to account for the inhomogeneity.

Powdered samples of Pr$_{2-x}$Bi$_{x}$Ru$_2$O$_7$ were synthesized by the solid state reaction method. The mixtures of Pr$_2$O$_3$, Bi$_2$O$_3$ and RuO$_2$ in proper molar ratios were pre-reacted at 850 $^\circ$C for 15 h in air and then ground and pressed into pellets. The pellets were sintered at 1000-1200 $^\circ$C in air with intermediate grindings. Powder X-ray diffraction measurements indicated single phase samples. Powder neutron diffraction data were collected on a $x\approx 1.0$ sample, using the BT1 diffractometer at NIST. Rietveld analysis confirmed the cubic pyrochlore structure and inductively coupled plasma-emission spectroscopy showed the Pr:Bi composition to be 1:0.94 corresponding to $x=0.97$. This sample was subsequently used for inelastic neutron scattering experiments.

\begin{figure}[t]
\includegraphics[width= 8cm]{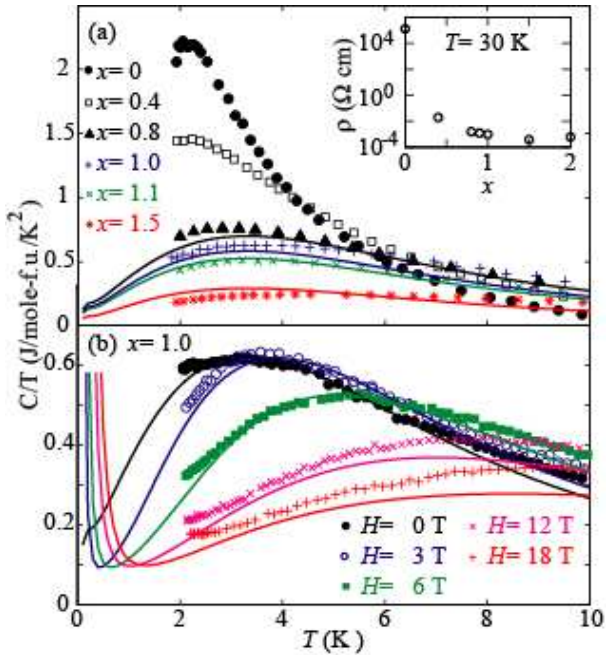}
\caption{Temperature dependence of the specific heats $C/T$ of Pr$_{2-x}$Bi$_{x}$Ru$_2$O$_7$ (a) for various $x$ and $H=0$ and (b) for various fields (H) for $x= 1.0$. The lattice contribution was subtracted as described in the text. The solid lines show $C/T$ calculated from the observed  neutron spectra and the assumption of a quenched level distribution (Eq.~(\ref{eq:Cp})). A $\gamma$ term of 0.04 J/mole/K$^2$ was added to the model in (b). Inset, electrical resestivity, $\rho$, of various $x$ at 30 K. \label{Cp}}
\end{figure}

Zero field heat capacity measurements were carried out on all members of the series using a commercial calorimeter. Field dependent heat capacity measurements employed a home made probe with a Si platform in a superconducting magnet. The electrical resistivity, $\rho$, of the polycrystalline samples was measured by the four-probe method. The overall temperature dependence of $\rho$ for different doping levels, $x$, are similar to those reported for Y$_{2-x}$Bi$_x$Ru$_2$O$_7$~\cite{YBiRu2O7_1,YBiRu2O7_2}. Pr$_2$Ru$_2$O$_7$ ($x=0$) is insulating while Bi$_2$Ru$_2$O$_7$ ($x=2$) is metallic (Fig.~\ref{Cp}). For intermediate values of $x$, $\rho$ decreases abruptly with $x$ indicating an insulator to metal transition for $x\approx$ 0.8.

\begin{figure}[t]
\includegraphics[width= 8cm]{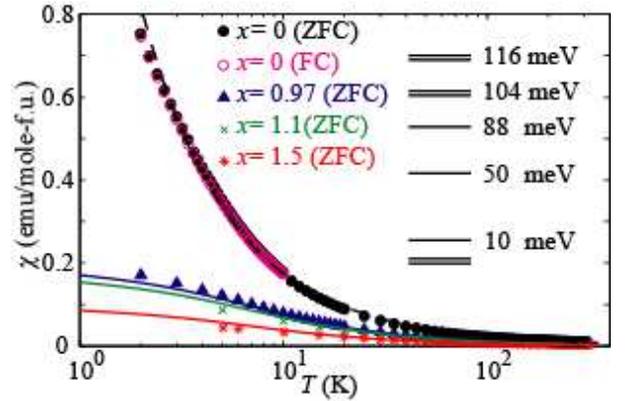}
\caption{Temperature dependence of the magnetic susceptibility $\chi$ of Pr$_{2-x}$Bi$_{x}$Ru$_2$O$_7$ for various $x$. The solid lines show the calculated magnetic susceptibility defined as $\chi(T)= \chi^{\prime}(T)+\chi_{CF}(T)$. Here $\chi^{\prime}(T)$ is given by Eq.~(\ref{eq:chi}) and $\chi_{CF}(T)$ is the contribution from the higher energy CF levels. The dashed line shows the calculated susceptibility for a doublet ground state. Inset, the Pr$^{3+}$ CF energy level scheme.\label{suscept}}
\end{figure}

Magnetic susceptibility data were collected on selected samples using a SQUID magnetometer. 
The insulating $x= 0$ sample shows abrupt development of hysteresis for $T<T_N=165$ K indicating N\'{e}el order plus spin canting from anisotropic exchange interactions~\cite{Pr2Ru2O7,Pr2Ru2O7_2}. While the $x=2$ sample behaves as a Pauli paramagnet~\cite{YBiRu2O7_2}, samples near the metal insulator transition for $x\approx 0.8$ have an increasing susceptibility upon cooling but fail to develop magnetic order down to $T=1.5$ K (Fig.~\ref{suscept}).

Fig.~\ref{Cp}(a) shows that metallic Pr$_{2-x}$Bi$_{x}$Ru$_2$O$_7$ samples close to the metal insulator transition have a strongly enhanced low$-T$ specific heat, $C(T)$. In this figure the lattice contributions have been subtracted using data from the $x=2$ sample corrected for unit cell mass through the Debye interpolation scheme. For $x=0.8$ and $T\approx 2.0$ K the specific heat ($C/T= 0.7$~J/mole-f.u./K$^2$) is more than three orders of magnitude greater than for a conventional metal. Figure~\ref{Cp}(b) shows that the enhanced specific heat is very sensitive to a magnetic field indicating a high density of magnetic states.

To determine their origin we performed inelastic neutron scattering experiments on the HET and IRIS spectrometers at the ISIS Facility
. On HET we used incident energies of E$_i$= 35 meV and 160 meV, which produced an elastic energy resolution of
 1.4 meV and 7 meV respectively. For the IRIS experiment disk choppers selected an incident energy spectrum from 1.3 meV to 4.6 meV pulsed at 25 Hz and a backscattering pyrolytic graphite analyser system with a 25 K Be filter selected a final energy of $E_f$= 1.847 meV for an elastic energy resolution of 17.5 $\mu$eV. Count rates were normalised to the incoherent scattering from a vanadium slab for the HET experiment, while IRIS data were normalized using coherent nuclear scattering from the sample. These procedures yield absolute measurements of
$
\tilde{I}(Q,\hbar\omega)= (\gamma r_0)^2 |\frac{g}{2}F(Q)|^2 2S(Q,\hbar\omega)
$
to an overall scale accuracy of 20 \%. Here $\gamma$= -1.913 and $g$ are spectroscopic g-factors of the neutron and magnetic ion respectively, $r_0$= 2.82 fm is the classical electron radius,  $F(Q)$ is the magnetic form factor for Pr~\cite{formfactor} and $S(Q,\hbar\omega)$ is the spherically averaged dynamic correlation function~\cite{lovesey,PrBiRu2O7_prb}.

To determine the relevant low energy degrees of freedom we used inelastic neutron scattering to map the crystal field (CF) level scheme for Pr$^{3+}$ in the sample. Five magnetic excitations were identified at 9.5(1) meV, 50(5) meV, 84.0(1) meV, 103.1(1) meV, and 115.7(2) meV~\cite{PrBiRu2O7_prb}. The number of excitations is consistent with a $J=4$ ion in $D_{3d}$ symmetry. The fact that excitations are observed to all 5 excited states implies a doublet ground-state. Further analysis of the spectrum indicates a ground state wave function of the form $|\pm\rangle=\alpha|\mp 4\rangle+\beta|\pm 2\rangle\pm \gamma|\mp 1\rangle$, where $\alpha=0.90(4)$, $\beta=-0.10(3)$, and $\gamma=0.42(7)$. CF excitations in the bismuth doped sample were broadened considerably as compared to pure Pr$_2$Ru$_2$O$_7$ where the excitations are almost resolution limited. The average doping induced half width at half maximum of 3.8 meV is a first indication of a distribution of CF environments.

\begin{figure}[t]
\includegraphics[width= 8cm]{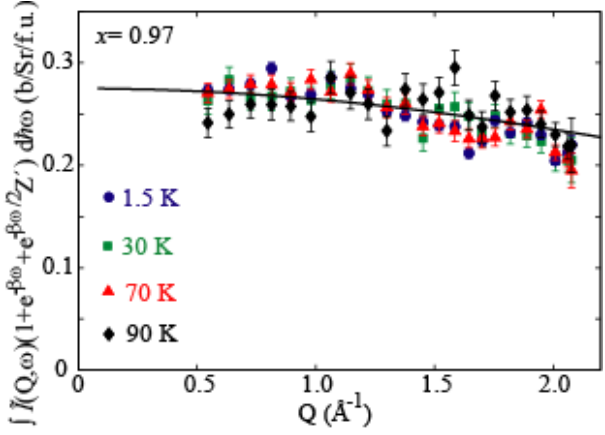}
\caption{$Q-$dependence of energy integrated inelastic neutron scattering intensity for $T=$ 1.5 K, 30 K, 70 K and 90 K. Data were obtained by integrating over $0.5 \leq \hbar\omega \leq 1.5$ meV. The solid line shows $|F(Q)|^2$ for Pr$^{3+}$ scaled to the data~\protect\cite{formfactor}.\label{Qdep}}
\end{figure}

\begin{figure}[t]
\includegraphics[width= 8cm]{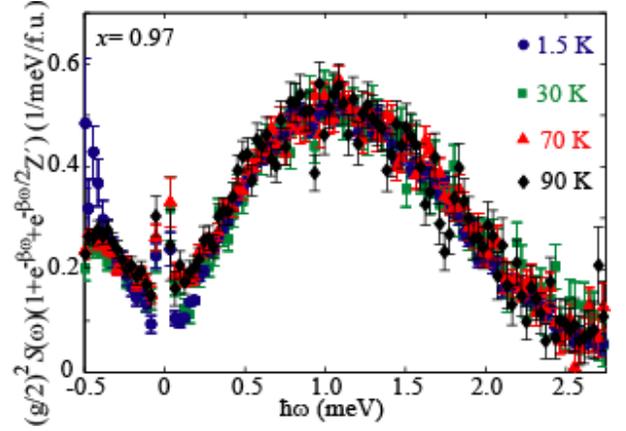}
\caption{Energy dependence of $Q-$averaged neutron scattering data scaled for proportionality to the level distribution function, $\rho(\Delta)$ of Eq.~\ref{eq:dist}. The data collapse indicates that $\rho(\Delta)$ is $T-$independent in the range probed. \label{hwdep}}
\end{figure}

The low energy magnetic response in the $x= 1.0$ sample features a broad and apparently dispersion-less peak centered at $\sim$1 meV energy transfer (Figs.~\ref{Qdep} and~\ref{hwdep}). The large spectral weight and the low energy scale suggests that these excitations are related to the enhanced low$-T$ specific heat of the material. The wave vector dependence of the intensity follows the squared single ion form factor of Pr$^{3+}$ (Fig.~\ref{Qdep}), indicating an absence of correlations between Pr sites and single ion physics. This is unusual for a dense Kondo (lattice) system, as in general competition between inter-site RKKY and on-site Kondo singlet formation produces short range antiferromagnetic correlations. The data also appear to be inconsistent with isolated Kondo singlets, which cannot scatter inelastically as $Q\rightarrow 0$~\cite{kondo_lat1, kondo_lat2, kondo_lat3}.  Most revealing, however, is that the spectrum, when appropriately scaled (Fig.~\ref{hwdep}), is utterly unaffected by heating to temperatures of order the N\'{e}el temperature for the parent insulator. This unusual result indicates that the finite width of the sample averaged spectrum results from some form of quenched inhomogeneity rather than relaxation effects.

We now pursue such an explanation for low energy physics in Pr$_{2-x}$Bi$_{x}$Ru$_2$O$_7$ more carefully. As Pr$^{3+}$ is a non-Kramers ion a doublet ground-state relies on $D_{3d}$ point group symmetry. Lowering of the CF symmetry with an amplitude that varies through the sample would split the doublet by varying amounts on different sites. Such quenched inhomogeneity would produce a $T$ independent spectrum of singlet-singlet transitions. As it is based on single site CF dynamics, this explanation links neutron scattering, specific heat, and susceptibility data in a parameter free fashion that we shall put to the test.

Denote the quenched distribution of singlet-singlet splitting by $\rho(\Delta)$ such that $N\rho(\Delta)d\Delta$ is the number of Pr ions with level splitting between $\Delta$ and $\Delta+d\Delta$ and N is the total number of formula units in the sample. Using expressions for the generalized susceptibility of the CF split doublet and the fluctuation dissipation theorem~\cite{lovesey} it is straightforward to show that the splitting distribution function is related to the sample averaged response functions as follows
\begin{eqnarray}
\rho(\hbar\omega)&=& \frac{\chi^{\prime\prime}(\hbar\omega)}{\pi (g\mu_B\alpha)^2}
\frac{1+e^{-\beta\hbar\omega}+e^{-\beta\hbar\omega/2}Z^{\prime}(\beta)}{1-e^{-\beta\hbar\omega}} \label{eq:dist}\\
&=&\frac{1}{\alpha^2}S(\hbar\omega)(1+e^{-\beta\hbar\omega}+e^{-\beta\hbar\omega/2}Z^{\prime}(\beta)).
\end{eqnarray}
Here $\beta= 1/k_BT$, $\chi^{\prime\prime}$ denotes the imaginary part of the generalized susceptibility, $g$= 0.8 is the spectroscopic g-factor for Pr$^{3+}$, $\alpha^2$ is the powder averaged squared singlet-singlet matrix element~\cite{powd_avg}, and $Z^{\prime}(\beta)= \sum_i^\prime n_i e^{-\beta E_i}$ is the partition sum excluding the low energy doublet. Fig.~\ref{hwdep} shows $(g/2)^2S(\hbar\omega)(1+e^{-\beta\hbar\omega}+e^{-\beta\hbar\omega/2}Z^{\prime}(\beta))$, which according to Eq.~(\ref{eq:dist}) and the underlying assumptions should be $T$ independent. The excellent data collapse provides strong evidence that the observed spectrum indeed results from inhomogeneous splitting of the Pr CF doublet. While the discrepancy at negative energies results from incomplete background subtraction, the differences in the data for $\hbar\omega\approx 0.1$~meV can be ascribed to incoherent inelastic phonon scattering on account of the approximately linear $T$ dependence.

Using perturbation theory and Eq.~(\ref{eq:dist}) the $T$ dependence of the specific heat and magnetic susceptibility can now be obtained directly from the neutron results for $\rho (\Delta)$ with {\em no} adjustable parameters.
\begin{eqnarray}
C_p(H,T)&=& k_B\int\frac{d\Omega_H}{4\pi}\int_0^{\infty}\rho(\Delta)(\beta\Delta(H_z))^2 \nonumber\\
 & & \times \frac{e^{\beta\Delta(H_z)}}{(1+e^{\beta\Delta(H_z)})^2} d\Delta
 \label{eq:Cp}
\end{eqnarray}
\begin{eqnarray}
\chi^{\prime}(T)&=& 2(g\mu_B\alpha)^2 \int_0^{\infty} \frac{\rho(\Delta)}{\Delta} \nonumber\\
 & & \times \frac{1-e^{-\beta\Delta}}{1+e^{-\beta\Delta}+e^{-\beta\Delta/2}Z^{\prime}(\beta)} d\Delta .
\label{eq:chi}
\end{eqnarray}
Here $\Delta(H_z)^2= (2g\mu_BH_z\langle +|J_z|+\rangle)^2+\Delta^2$ and $H_z$ is the component of the magnetic field along the local 3-fold axis. Note that Eq.~(\ref{eq:chi}) becomes the exact Kramers-Kronig relation when $\rho (\Delta)$ from Eq.~(\ref{eq:dist}) is evaluated at the running temperature. However, the assumption of a quenched distribution backed by Fig.~\ref{hwdep} allows calculation of the full temperature dependence of $\chi^{\prime}(T)$ based on a single neutron spectrum. To minimize contributions from phonon scattering we used $T= 1.5$ K data to derive $\rho (\Delta)$. Note however that indistinguishable results are obtained using the average of all temperatures.

The lines in Fig.~\ref{Cp} were calculated from Eq.~(\ref{eq:Cp}) neglecting any variation of $\rho (\Delta)$ with Pr concentration. We added the nuclear specific heat as calculated from the natural isotope distribution of the sample~\cite{Csch}. In the bottom frame we added a common best fit Sommerfeld term, $\gamma_{\rm Ru}=$40 mJ/mole/K$^2$, to account for contributions from the metallic state. This value is not dissimilar from  $\gamma_{\rm Ru}=$72 mJ/mole/K$^2$ for Pauli paramagnetic $\rm YBiRu_2O_7$. The magnetic susceptibility was calculated as $\chi(T)= \chi^{\prime}(T)+\chi_{CF}(T)$, where $\chi^{\prime}(T)$ is given by Eq.~(\ref{eq:chi}) and $\chi_{CF}(T)$ accounts for contribution from higher energy CF levels~\cite{chi_calc}. The discrepancy between the lines and the data points is of order the accuracy in the overall normalization of the specific heat, susceptibility, and neutron data. As there are {\em no} adjustable parameters this supports the inference that Pr ions in $\rm Pr_{2-x}Bi_{x}Ru_2O_7$ form a distributed two level system.

The measured distribution function provides important clues to the origin of level splitting in $\rm Pr_{2-x}Bi_{x}Ru_2O_7$. The absence of $T$ dependence to  $\rho (\Delta)$ from 1.5 K to 90 K (see Fig.~\ref{hwdep}) all but rules out dynamic phenomena. Fluctuations slow enough to be quasi-static on the time scale of the $\approx 1$ meV singlet-singlet transitions would be thermally activated in that temperature range. So rather than reflecting changes with time for a single Pr ion, $\rho (\Delta)$ represents a spatial distribution of Pr environments. Because Eq.~(\ref{eq:Cp}) accounts for finite field data too, magneto-static disorder from high T Ru spin freezing seems unlikely though this deserves further examination through $\mu$SR measurements. To produce a continuous distribution rather than a finite set of levels the length scale associated with this disorder must greatly exceed the lattice spacing. Possible mechanisms include a low density of extended defects or a density wave which generates a continuum of local environments. The natural culprit would be Bi doping however this would appear to be at odds with the fact that a single distribution function measured for $x=0.97$  accounts for all specific heat data for $x>0.8$. There are in fact indications of disorder for both end members of the series. While the high $T$ specific heat anomaly for Pr$_2$Ru$_2$O$_7$ indicates quasi-long-range N\'{e}el order below $T_N({\rm Ru})=165$ K, the phase transition is accompanied by a difference in the field cooled and zero field cooled susceptibility. Furthermore, with a residual resistivity of 1.5 m$\Omega$cm the metallic end member, Bi$_2$Ru$_2$O$_7$ appears to be severely disordered. These results suggest that a common low density of large defects principally controlled by the ruthenium sub-lattice may influence Pr$_{2-x}$Bi$_{x}$Ru$_2$O$_7$ at least through the metallic part of the phase diagram.

Pyrochlore systems are indeed known to suffer from various types of clandestine disorder. $\rm Y_2Mo_2O_7$ was for example recently shown to have significant disorder on the Mo sublattice~\cite{booth}. Virtually undetectable through diffraction this quenched disorder nonetheless causes spin freezing for $T<T_g=22.5$ K. Just as non-Kramers doublets are particularly sensitive to disorder, so are highly frustrated magnets so that modest levels of disorder that might not affect more conventional magnetic materials can have qualitative effects in these systems. Materials such as Pr$_{2-x}$Bi$_{x}$Ru$_2$O$_7$ and Pr$_2$Ir$_2$O$_7$~\cite{yanagishima} that are \emph{both} magnetically frustrated {\em and} contain a non-kramers ion are particularly susceptible to these effects.

Non-Kramers magnetic ions include Pr$^{3+}$, Pm$^{3+}$, Sm$^{2+}$, Eu$^{3+}$, Tb$^{3+}$, Ho$^{3+}$ and Tm$^{3+}$ from the rare earth series and U$^{4+}$ from the actinides. There are a number of materials of current interest where distributed lifting of a non-Kramers degeneracies might account for unusual low$-T$ anomalies. These include nominally stoichiometric materials such as Tb$_2$Ti$_2$O$_7$~\cite{gardner}, and  Ho$_2$Ti$_2$O$_7$~\cite{harris}. There may also be specific interesting effects of doping non-kramers systems as in LiHo$_x$Y$_{1-x}$F$_4$~\cite{brooke} and Tb$_x$Y$_{2-x}$Ti$_2$O$_7$~\cite{keren} where symmetry breaking impurities may act as a random transverse field.

It is a pleasure to acknowledge helpful discussions with O. Tchernyshyov, B. D. Rainford and A. Migliori for his assistance. Work at JHU and ISIS was supported by DoE through DE-FG02-02ER45983. Work at Rutgers was supported by NSF through DMR-0103858. KHK was partially supported by KOSEF through CSCMR.

\end{document}